\documentclass[twocolumn,prd,floatfix,amsmath,nofootinbib,amssymb,floatfix]{revtex4}
\usepackage{graphicx,color,dcolumn,booktabs,bm}
\usepackage{longtable,lscape}
\usepackage{txfonts}
\usepackage{overpic}
\usepackage{amssymb}
\usepackage{indentfirst}
\usepackage{feynmf}
\usepackage{slashed}
\usepackage{cases}
\usepackage{color}
\usepackage{multirow}
\usepackage{longtable}
\usepackage{ulem}
\usepackage{graphicx,color,dcolumn,booktabs,bm}
\usepackage[colorlinks,
            citecolor=blue,
            anchorcolor=red,
            menucolor=red,
            linkcolor=red,
            filecolor=red,
            runcolor=red,
            urlcolor=blue,
            frenchlinks=red]{hyperref}
\usepackage{epstopdf}
\begin{document}

\title{Resolving the low mass puzzle of $\Lambda_c(2940)^+$}

\author{Si-Qiang Luo$^{1,3}$}\email{luosq15@lzu.edu.cn}
\author{Bing Chen$^{2,3}$}\email{chenbing@ahstu.edu.cn}
\author{Zhan-Wei Liu$^{1,3}$}\email{liuzhanwei@lzu.edu.cn}
\author{Xiang Liu$^{1,3}$\footnote{Corresponding author}}\email{xiangliu@lzu.edu.cn}

\affiliation
{
$^1$School of Physical Science and Technology, Lanzhou University, Lanzhou 730000, China\\
$^2$School of Electrical and Electronic Engineering, Anhui Science and Technology University, Fengyang 233100, China\\
$^3$Research Center for Hadron and CSR Physics, Lanzhou University and Institute of Modern Physics of CAS, Lanzhou 730000, China
}

\begin{abstract}

For the long standing low mass puzzle of $\Lambda_c(2940)^+$, we propose an unquenched picture.
Our calculation explicitly shows that the mass of the $\Lambda_c(2P,3/2^-)$ state can be lowered down to be consistent with
the experimental data of $\Lambda_c(2940)^+$ by introducing the $D^*N$ channel contribution. It means that the low mass puzzle of $\Lambda_c(2940)^+$ can be solved. What is more important is that we predict a mass inversion relation for the $2P$ $\Lambda_{c}^+$ states, i.e., the $\Lambda_c(2P,1/2^-)$ state is higher than the $\Lambda_c(2P,3/2^-)$, which is totally different from the result of conventional quenched quark model. It provides a criterion to test such an unquenched scenario for $\Lambda_c(2940)^+$. We expect the future experimental progress from the LHCb and Belle II.

%\pacs{12.39.Jh, 13.30.Eg, 14.20.Lq}

\end{abstract}

\maketitle

\noindent {\it{Introduction}:--}How to quantitatively depict the non-perturbative behavior of quantum chromodynamics (QCD) is always a big problem in particle physics. Studying on the hadron spectroscopy is an effective approach to deepen our understanding of such problem. In the past decades, hadron family has become more and more abundant, which can be reflected by {the increasing number of these observed hadronic states collected into} the ``Review of Particle Physics'' (RPP)~\cite{Tanabashi:2018oca} compiled by the Particle Data Group (PDG) due to the efforts from the particle physicists. A recent typical example is these observed charmoniumlike $XYZ$ states which have stimulated the extensive exploration of multiquark hadronic matter. It not only has become a hot issue of hadron physics but also provided valuable hints to further probe into the non-perturbative behavior of QCD (see Ref.~\cite{Chen:2016qju,Liu:2019zoy} for a comprehensive review).

Charmed baryon family occupies a special position in the hadron spectroscopy. With the efforts of the CLEO, BaBar, CDF, Belle, and recent LHCb collaborations, the $\Lambda_c(2286)^+$, $\Lambda_c(2595)^+$, $\Lambda_c(2625)^+$, $\Lambda_c(2760)^+$, $\Lambda_c(2860)^+$, $\Lambda_c(2880)^+$, $\Lambda_c(2940)^+$, $\Sigma_c(2455)^{0,+,++}$, $\Sigma_c(2520)^{0,+,++}$, and $\Sigma_c(2880)^{0,+,++}$ have been established by experiments (see Sec.~2.5 of Ref.~\cite{Chen:2016spr}). As indicted in Refs.~\cite{Ebert:2011kk,Chen:2016iyi,Chen:2017aqm}, these states can be categorized into the charmed baryon family without doubt, except for $\Lambda_c(2940)^+$.

The $\Lambda_c(2940)^+$ was first observed in the $D^0p$ mass spectrum by the BaBar Collaboration~\cite{Aubert:2006sp}, and later, confirmed by Belle in decay mode $\Sigma_c(2455)\pi$~\cite{Abe:2006rz}. More importantly, in 2017, the LHCb further measured $\Lambda_c(2940)^+$ as a $P$-wave state with $J^P=3/2^-$~\cite{Aaij:2017vbw}. Now, PDG listed the mass and decay width of $\Lambda_c(2940)^+$ as $M=2939.6^{+1.3}_{-1.5}$ MeV and $\Gamma=20^{+6}_{-5}$ MeV~\cite{Tanabashi:2018oca}. There exists low mass puzzle for the $\Lambda_c(2940)^+$, which results in the difficulty to arrange $\Lambda_c(2940)^+$ under the framework of charmed baryon \cite{Ebert:2011kk,Chen:2016iyi,Aubert:2006sp,Aaij:2017vbw,Capstick:1986bm,Chen:2014nyo,Cheng:2017ove}. Due to this reason, the exotic hadronic molecular configuration to $\Lambda_c(2940)^+$ was proposed~ \cite{He:2006is}. However, we need to exhaust different possibilities under the conventional framework before confirming the existence of exotic state. Along this line, it is obvious that we are not satisfied with the present solution to the low mass puzzle of $\Lambda_c(2940)^+$ when looking back on the present research status of $\Lambda_c(2940)^+$. New idea must emerge for solving this low mass puzzle.

\begin{figure}[htbp]
\centering
\includegraphics[width=8.6cm,keepaspectratio]{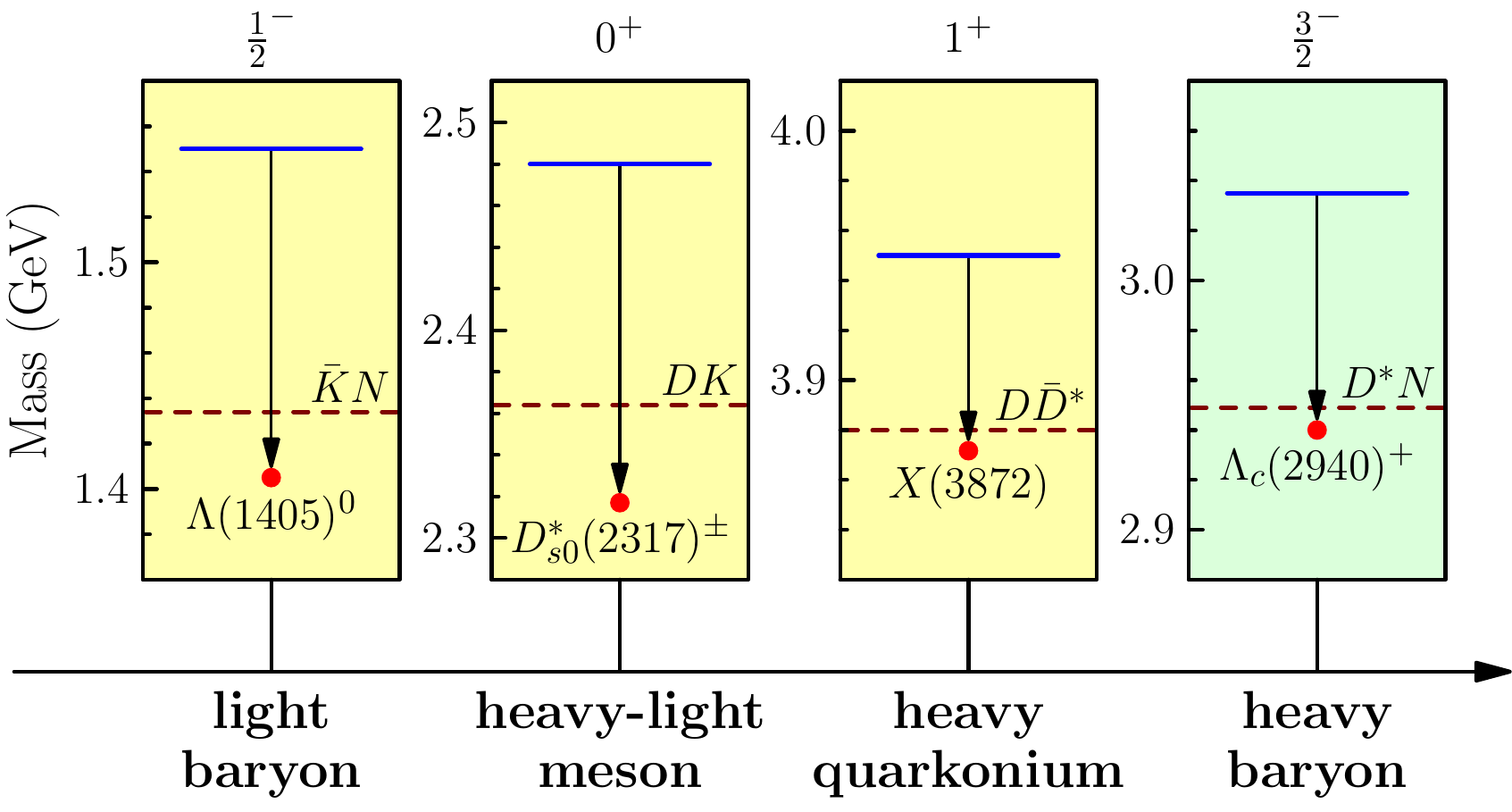}
\caption{The comparisons of the masses in experiments and for the corresponding undressed state within the conventional quark model. The dashed lines represent the thresholds. The dots are measured results, and the blue solid lines are theoretical masses which obtained from~\cite{Godfrey:1985xj,Capstick:1986bm}. The $D_{s1}^\prime(2460)^{\pm}$ ($J^P=1^+$) is close to the $D^*K$ threshold and nearly 90~MeV below the quenched one, and it is also a heavy-light meson as $D^*_{s0}(2317)^\pm$ and thus not shown. {For $\Lambda_c(2940)^+$, LHCb suggested that the most likely spin-parity $J^P$ is $3/2^-$ \cite{Aaij:2017vbw}. Thus, we adopted $J^P=3/2^-$ assignment to $\Lambda_c(2940)^+$ in our discussion of $\Lambda_c(2940)^+$ in this work.}}
\label{nearbythresholdstates1}
\end{figure}

When checking the whole observed hadrons, we notice four states $\Lambda(1405)^0$, $D^*_{s0}(2317)^\pm$, $D^\prime_{s1}(2460)^\pm$, and $X(3872)$, which have been established in experiments. If further comparing $\Lambda_c(2940)^+$ with these four states, we find the similarities: 1) there exists low mass problem, i.e., they are are about $100$ MeV lower than the corresponding theoretical results from ordinary (quenched) quark model calculation; 2) they are close to some $s$-wave channel thresholds as shown in the figure%\footnote{\textcolor{red}{The lowercase $s$-, $p$-, and $d$-waves represent the orbital angular momentums for the two hadrons while the capital $S$-, $P$-, and $D$-waves are those among quenched quarks in hadrons.}}
; 3) especially they have $P$-wave quantum number. In Fig.~\ref{nearbythresholdstates1}, we illustrate these common features.

In fact, we may draw inspiration from the research progress around $\Lambda(1405)^0$, $D^*_{s0}(2317)^\pm$, $D^\prime_{s1}(2460)^\pm$, and $X(3872)$. For understanding low mass problems existing in these states, the unquenched quark model by including coupled-channel effect was developed, which
was applied to explain why the masses of the corresponding physical states can be lowered down to be consistent with experimental data (see Refs.~ \cite{SilvestreBrac:1991pw,vanBeveren:2003kd,Kalashnikova:2005ui,Li:2009ad,Danilkin:2010cc} for example). Due to the similarities illustrated above, we naturally conjecture whether unquenched picture can happen for discussed $\Lambda_c(2940)^+$.

In this work, we construct an unquenched picture to test such a scenario. Our calculated results explicitly show that the low mass puzzle of $\Lambda_c(2940)^+$ can be solved, which provides a unique view point to decode the nature of $\Lambda_c(2940)^+$ without including so called exotic state assignment to $\Lambda_c(2940)^+$. Success of solving the low mass puzzle of $\Lambda_c(2940)^+$ makes that $\Lambda_c(2940)^+$ becomes the first typical example
affected by the unquenched effect in the heavy baryon family. What is more important is that group of $\Lambda_c(2940)^+$ with  $\Lambda(1405)^0$, $D^*_{s0}(2317)^\pm$, $D^\prime_{s1}(2460)^\pm$, and $X(3872)$ constructs a complete chain. Until now, the unquenched effect can be seen
in different types of hadronic system (from light baryon to charmed-strange meson containing heavy-light quarks, $c\bar{c}$ double-heavy meson system, and to charmed baryon with heavy-light quarks), which is a fantastic phenomenon.

Besides solving the low mass puzzle of $\Lambda_c(2940)^+$, we predict a mass inversion for the $2P$ $\Lambda_c^+$ states,
i.e., the mass of $2P$ $1/2^-$ $\Lambda_c^+$ state is expected to be larger than that of $\Lambda_c(2940)^+$ under the unquenched picture.
We should emphasize that there must exist such mass inversion relation for the $2P$ $\Lambda_c^+$ states if the unquenched effect plays an important role to $\Lambda_c(2940)^+$\footnote{In the quenched picture, $1/2^-$ state must be lower than $3/2^-$ states for the $2P$ $\Lambda_c^+$ states.}.
Since the predicted $2P$ $1/2^-$ $\Lambda_c^+$ state is still missing, searching for this missing state becomes a crucial point to test the unquenched scenario for $\Lambda_c(2940)$. It will be a good opportunity for the future experimental study at LHCb and Belle II.

\noindent {\it{An unquenched picture for $\Lambda_c(2940)^+$}:--}
Before introducing the unquenched picture for $\Lambda_c(2940)^+$, we firstly mention what is the low mass puzzle of $\Lambda_c(2940)^+$. According to the calculations of quenched quark models~\cite{Chen:2016iyi,Chen:2014nyo,Capstick:1986bm,Ebert:2011kk} and Regge trajectory analysis~\cite{Cheng:2017ove}, the mass of $\Lambda_c(2P)$ state with $J^P=3/2^-$, which is tentatively named as $\Lambda_c(2P,3/2^-)$ for convenience of later discussion, should be in the range $3000\sim 3040$ MeV, which is $60\sim 100$ MeV larger than the measured resonance parameter of $\Lambda_c(2940)^+$. This phenomenon results in the confusion for its nature in past years. Making comparison with the $D^*N$ threshold, we notice that $\Lambda_c(2P,3/2^-)$ may couple with this $D^*N$ channel via $s$-wave interaction. In fact, the situation of $\Lambda_c(2P,3/2^-)$ is similar to that of several typical states like $\Lambda(1405)^0$, $D_{s0}^*(2317)^{\pm}$, $D_{s1}^\prime(2460)^{\pm}$, and $X(3872)$, where the masses of corresponding bare states are larger than the corresponding observed values and there exist $s$-wave couplings between the typical states with the concrete thresholds.

{In our calculation, we only select the $D^*N$ channel contribution to the discussed $\Lambda_c(2P,3/2^-)$. This treatment is due to suggestion by Geiger and Isgur in Ref.~\cite{Geiger:1989yc,Isgur:1998kr}. Usually, all possible hadronic channels coupled with a bare state should be included. But, this consideration makes the calculation become impractical~\cite{Geiger:1989yc}, where the trivial and nontrivial coupled channel cases should be distinguished by different treatments \cite{Geiger:1989yc,Isgur:1998kr}. Isgur indicated that {the long-range coupled channel effects from the nonperturbative quark loops can be absorbed by the string tension} while the $q\bar{q}$ pair creation at short distances just changes the running coupling constant $\alpha_s$~\cite{Isgur:1998kr}. Thus, in most cases, the trivial coupled channel effect can be renormalized in the parameters $\alpha_s$ and $b$ in the quenched quark model. It also naturally explains why most of the observed meson and baryon states can be depicted in the quenched quark models~\cite{Godfrey:1985xj,Capstick:1986bm}.
Isgur further pointed out that the nontrivial coupled channel effect can not be treated as the adiabatic approximation when a resonance state strongly couples to nearby $s$-wave threshold~\cite{Isgur:1998kr}. It is the cases of $\Lambda(1405)$, $D^\ast_s(2317)^\pm$, $D^\prime_s(2460)^\pm$, and $X(3872)$ states since they couple strongly with the nearby $NK$, $DK$, $D^\ast{K}$, and $\bar{D}^\ast{D}$ thresholds, respectively. In this work, the $\Lambda_c(2940)^+$ has been verified as the first known heavy baryon state which should be considered the nontrivial coupled channel effect seriously.
}

For reflecting the contribution from the $D^*N$ channel, we need to write out the so called {\it full} Hamiltonian of the physical $\Lambda_c(2P,3/2^-)$~\cite{Guo:2017jvc,Tornqvist:1984fy, Barnes:2007xu, Pennington:2007xr, Tornqvist:1995kr, Kalashnikova:2005ui, Zhou:2011sp}
\begin{equation}\label{full_h}
\hat{H}=\hat{H}_0+\hat{H}_I+\hat{H}_{D^*N},
\end{equation}
where $\hat{H}_0$ depicts the discrete mass spectrum of the bare charmed baryon, which has expression
\begin{equation*}%\label{h0}
\hat{H}_0=\sum\limits_{i=1}^{3}\left(m_i+\frac{p_i^2}{2m_i}\right)+\sum\limits_{i<j}{\bf F}_i\cdot{\bf F}_j\left[\left(\frac{\alpha_s}{r_{ij}}-\frac{3}{4}br_{ij}-\frac{3}{4}C\right)+V^{\rm spin}_{ij}\right].
\end{equation*}
Here, the parameters $\alpha_c$, $b$, and $C$ represent the strength of the color Coulomb potential, the strength of linear confinement, and a mass-renormalized constant, respectively. The spin-dependent interactions, $V^{\rm spin}_{ij}$, contain the spin-spin contact hyperfine interaction, the tensor interaction, the spin-orbit interaction of color-magnetic piece, and the Thomas precession term (see Refs.~\cite{Godfrey:1985xj,Capstick:1986bm} for more details). The color factor $\langle{\bf F}_i\cdot{\bf F}_j\rangle$ is taken as $-2/3$ for the baryon system (the meson system,  $\langle{\bf F}_i\cdot{\bf F}_j\rangle=-4/3$). In our calculation, the masses of $u/d$ and $c$ quarks are taken as 0.370 GeV and 1.88 GeV, respectively. For the baryons, the parameters $\alpha_s$, $b$, and $\sigma$ (where the $\sigma$ is a parameter in contact term, and one can refer from~\cite{Close:2005se}) are taken as 0.554, 0.120 GeV$^2$, and 1.60 GeV, respectively. In our work, we also reproduce the masses of light and charmed mesons since their wave functions shall be used in our unquenched calculation. The values of $\alpha_s$, $b$, and $\sigma$ for these meson systems are taken as 0.561, 0.142 GeV$^2$, and 1.08 GeV, respectively. Finally, the constant $C$ are fixed as $C_{udc}=-0.630$ GeV, $C_N=-0.746$ GeV, $C_\pi=-0.655$ GeV, and $C_D=-0.700$ GeV for the different hadron systems.

With the Hamiltonian $\hat{H}_0$ and parameters presented above, the Gaussian Expansion Method~\cite{Hiyama:2003cu} is adopted to solve the Schr\"{o}dinger equations. The bare mass of $\Lambda_c(2P,3/2^-)$ is obtained to be 3012 MeV, which is about 70 MeV larger than the mass of the discussed $\Lambda_c(2940)^+$. In fact, basing on this Hamiltonian $\hat{H}_0$ the masses of $\Lambda_c(1S,1/2^+)$, $\Lambda_c(2S,1/2^+)$, $\Lambda_c(1P,1/2^-)$, $\Lambda_c(1P,3/2^-)$, $\Lambda_c(1D,3/2^+)$, and $\Lambda_c(1D,5/2^+)$ can be given as 2287, 2779, 2595, 2617, 2856, and 2863 (in units of MeV), which may correspond to  these  well-established $\Lambda_c(2286)^+$, $\Lambda_c(2765)^+$, $\Lambda_c(2595)^+$, $\Lambda_c(2625)^+$, $\Lambda_c(2860)^+$, and $\Lambda_c(2880)^+$, respectively. This fact shows that Hamiltonian $\hat{H}_0$ works well to reproduce most of charmed baryons.

\begin{figure}[htbp]
\centering
\includegraphics[width=8.6cm,keepaspectratio]{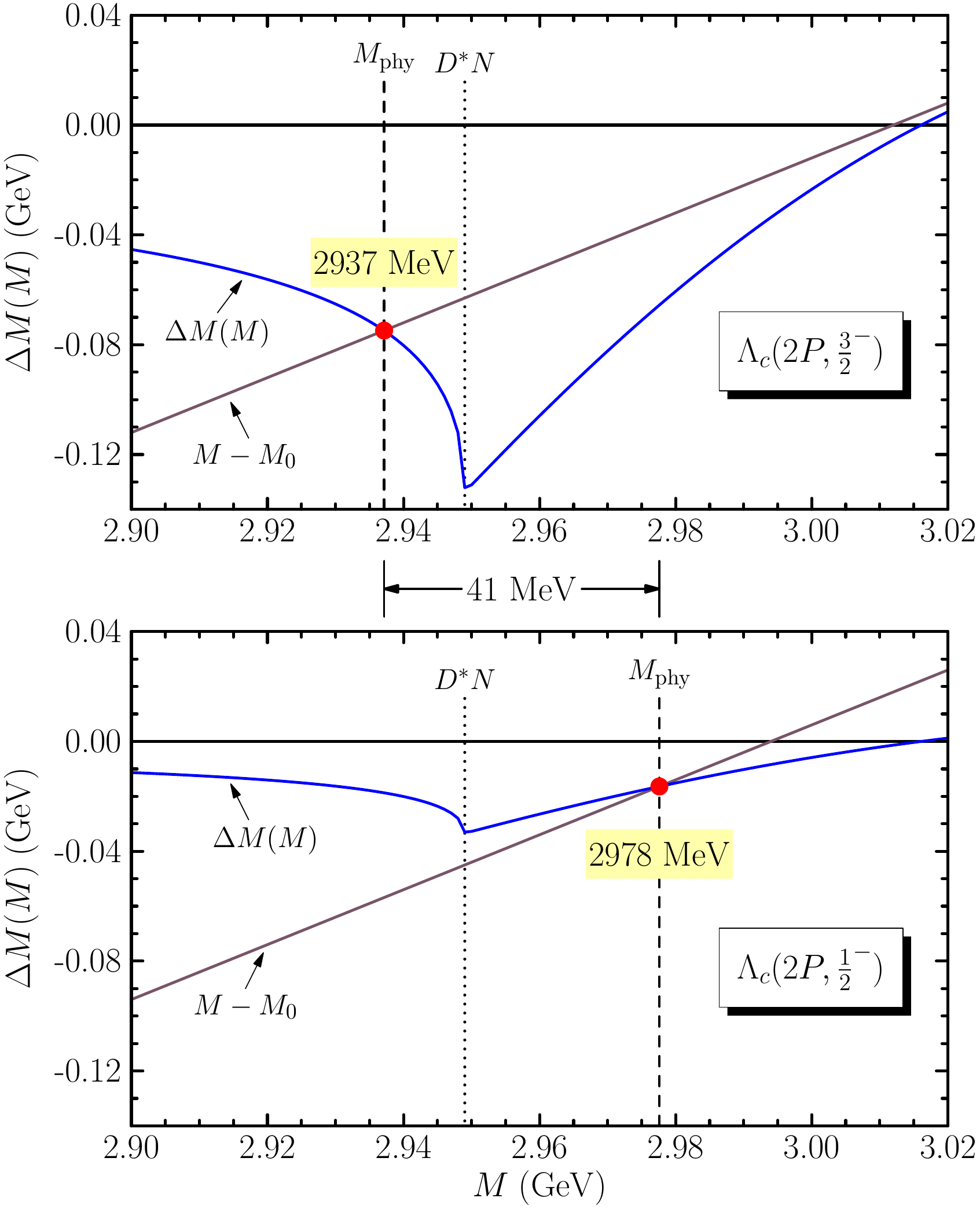}
\caption{The dependence of functions $M-M_0$ and $\Delta M(M)$ on $M$ for $\Lambda_c(2P,3/2^-)$ (up) and $\Lambda_c(2P,1/2^-)$ (down). Here, the $M_{\rm phy}$ values correspond to the red points of the intersections of two lines are physical masses of the $\Lambda_c(3/2^-,2P)$ (up) and $\Lambda_c(1/2^-,2P)$ (down) states. The gap between the two subgraphs represents that the physical mass of $\Lambda_c(2P,1/2^-)$ is 41 MeV higher than that of $\Lambda_c(2P,3/2^-)$.}
\label{chlambdac2p}
\end{figure}

In Eq.~(\ref{full_h}), the Hamiltonian $\hat{H}_I$ describes the interaction between the bare state and the $D^*N$ channel, which is responsible for the dress of the bare state. In this work, we employ $\hat{H}_I=g\int{d^3x}\bar{\psi}(x)\psi(x)$ inspired by the quark-pair-creation (QPC) model. In the non-relativistic limit, this $\hat{H}_I$ can be replaced by~\cite{Chen:2016iyi}
\begin{equation}\label{toperator}
\begin{split}
\hat{H}_I=&-3\gamma\sum_{m}\langle 1,m;1,-m|0,0\rangle\int{\rm d}^3{\bf p}_\mu{\rm d}^3{\bf p}_\nu\delta({\bf p}_\mu+{\bf p}_\nu)\\
                  &\times\mathcal{Y}_1^m\left(\frac{{\bf p}_\mu-{\bf p}_\nu}{2}\right)\omega^{(\mu,\nu)}\phi^{(\mu,\nu)}\chi_{-m}^{(\mu,\nu)}b^\dagger_\mu({\bf p}_\mu)d^\dagger_\nu({\bf p}_\nu),
\end{split}
\end{equation}
where the $\omega$, $\phi$, $\chi$ and $\mathcal{Y}^m_1$ are the color, flavour, spin, and orbit functions of the quark pair, respectively. The $b_{\mu}^\dagger$ and $d_{\nu}^\dagger$ are quark and antiquark creation operators, respectively. The dimensionless parameter $\gamma$ describes the strength of the quark-antiquark pair created from the vacuum, which is fixed as 9.45 by the total decay width of $\Sigma_c(2520)$~\cite{Tanabashi:2018oca}.

When the $D^*N$ channel effect is taken into account, the physical state $\Lambda_c(2P,3/2^-)$, which contains a significant continuum component of $D^*{N}$ other than the $udc$ component, can be represented as
\begin{equation*}
\left|\Lambda_c(2P,3/2^-)\right\rangle=c_0\,\left|\Lambda^{\rm bare}_c(2P,3/2^-)\right\rangle+\int{\textrm{d}^3\textbf{q}}\,\chi(\textbf{q})\,\left|D^*{N},\textbf{q}\right\rangle.
\end{equation*}
Here, the $c_0$ denotes the probability amplitude of the $udc$ core in $\Lambda_c(2940)^+$, and the $\chi(\textbf{q})$ is the wave function of the $|D^*{N}\rangle$ channel.
Finally, the physical mass $M_{\rm phy}$ for $\Lambda_c(2P,3/2^-)$ affected by the $D^*{N}$ channel can be obtained from the following equations for $M$
\begin{equation}\label{coupledchannelmass1}
M-M_0-\Delta M(M)=0,
\end{equation}
\begin{equation}\label{coupledchannelmass2}
\Delta M(M)={\rm Re}\;\int_0^\infty {\rm d}q\;q^2\frac{\left|{\cal M}^{\Lambda_c^{\rm bare}(2P,3/2^-)\to D^\ast N}(q)\right|^2}{M-E_{D^\ast N}(\textbf{q})},
\end{equation}
where the $M_0=3012$ MeV is the bare mass of $\Lambda_c(2P,3/2^-)$, which is already mentioned above.
The transition amplitude ${\cal M}^{\Lambda_c^{\rm bare}(2P,3/2^-)\to D^* N}(q)$ can be calculated by the QPC model, i.e., ${\cal M}^{\Lambda_c^{\rm bare}(2P,3/2^-)\to D^* N}(q)=\delta({\bf P}_{D^*}+{\bf P}_N)\left\langle D^*{N},q\left|\hat{H}_I\right|\Lambda_c^{\rm bare}(2P,3/2^+)\right \rangle$. {The $E_{D^* N}(\textbf{q})=\sqrt{M_{D^*}^2+\textbf{q}^2}+\sqrt{M_{N}^2+\textbf{q}^2}$ is the energy of $|D^*{N},\textbf{q}\rangle$ component which is from the Hamiltonian $\hat{H}_{D^*N}$ in Eq.~(\ref{full_h}) .} {{$c_0$ can be determined by following expression 
\begin{equation}
|c_0|^2=\left(1-\left.\frac{\partial {\rm Re}\;\Delta M(M)}{\partial M}\right|_{M=M_{\rm phy}}\right)^{-1},
\end{equation}
where the $M_{\rm phy}$ is the solution of Eqs.~(\ref{coupledchannelmass1})-(\ref{coupledchannelmass2}) for $M$.
}}

With the above preparation of quantitatively constructing unquenched model, we illustrate how to extract the physical mass of $\Lambda_c(2P,3/2^-)$. As shown in the left diagram of Fig. \ref{chlambdac2p}, we plot the functions $M-M_0$ and $\Delta M(M)$ dependent on $M$. A vivid cusplike behavior near the $D^*N$ threshold is exhibited. The intersection of two curves corresponds to the physical mass $M_{\rm phy}$ of $\Lambda_c(2P,3/2^-)$. Our calculation explicitly reveals how the $D^*N$ channel contribution lowers the bare mass 3012 MeV to the physical mass 2937 MeV, which is consistent with experimental data of the observed $\Lambda_c(2940)^+$. Thus, the low mass puzzle of $\Lambda_c(2940)^+$ can be solved well in this unquenched picture.

When checking the resonance parameters of the observed $\Lambda_c(2940)^+$, we notice that $\Lambda_c(2940)^+$ is a narrow state with width 20~ MeV~\cite{Aubert:2006sp,Abe:2006rz,Aaij:2017vbw,Tanabashi:2018oca}, which can be also understood by a simple analysis mentioned below.
There were theoretical calculations for the strong decay behavior of the $\Lambda_c(2P,3/2^-)$ state in the quenched quark model, by which the total width of $\Lambda_c(2P,3/2^-)$ is predicted at about 380~MeV. Similar theoretical results were also obtained in Refs.~\cite{Lu:2018utx,Lu:2019rtg}. It is obvious that these calculations are not consistent with experimental measurements of $\Lambda_c(2940)^+$ as a narrow state. In this work, our unquenched calculation indicates the $\Lambda_c(2P,3/2^-)$ state is below the $D^*N$ threshold, which means that the $D^*N$ decay channel is forbidden kinematically. Thus, the main contribution to the width of $\Lambda_c(2P,3/2^-)$ has to disappear, which makes $\Lambda_c(2P,3/2^-)$ possible as a narrow state. Especially, since the mass of $\Lambda_c(2P,3/2^-)$ is lowered down to be consistent with $\Lambda_c(2940)^+$, the phase space of $\Lambda_c(2P,3/2^-)$ becomes smaller, which may directly result in $\Lambda_c(2P,3/2^-)$ being a narrow state. According to this analysis, assigning the $\Lambda_c(2940)^+$ as a $\Lambda_c(2P,3/2^-)$ is suitable in the unquenched framework.

Besides focusing on $\Lambda_c(2940)^+$, in this letter we also study its partner $\Lambda_c(2P,1/2^-)$.
The results shown in Fig.~\ref{chlambdac2p} indicate that the mass of $\Lambda_c(2P,1/2^-)$ should locate at 2978~MeV, where we adopt the same analysis approach and input parameters for $\Lambda_c(2P,3/2^-)$. The mass shift due to the $D^*N$ channel contribution is not enough to make the mass of $\Lambda_c(2P,1/2^-)$ shift down below the $D^*N$ threshold, which is different from the case of $\Lambda_c(2P,3/2^-)$.
It leads to an interesting phenomenon happening in charmed baryon family, i.e., we find a mass inversion relation of the $2P$ $\Lambda_c^+$ states. If this unquenched effect due to the $D^*N$ channel plays roles for the $\Lambda_c(2P,3/2^-)$ and $\Lambda_c(2P,1/2^-)$, this mass inversion relation cannot be avoided. When checking the PDG data for heavy {baryons}, we cannot find similar situation. It means that it will be the first time to predict and find novel phenomenon of mass inversion relation in heavy flavor {baryon} sectors. In fact, this predicted mass inversion phenomenon can be directly applied to seriously test our unquenched scenario for the $\Lambda_c(2P)$ states.

\noindent {\it{Summary}:--}Since the observation of $\Lambda_c(2940)^+$, the low mass puzzle has been there. We have no any reason to ignore this problem since it is a key point to reveal its nature. Noticing the similarities between $\Lambda_c(2940)^+$ and several
typical states like $\Lambda(1405)^0$, $D_{s0}^*(2317)^{\pm}$, $D_{s1}^\prime(2460)^{\pm}$, and $X(3872)$, in this letter we propose an unquenched picture to study $\Lambda_c(2940)^+$. Due to the $D^*N$ channel contribution, which couples with $\Lambda_c(2P,3/2^-)$ in $s$-wave, the mass of
$\Lambda_c(2P,3/2^-)$ can be lowered down to 2937~MeV. Our calculation reproduces the mass of $\Lambda_c(2940)^+$, which
provides a direct solution to solve the low mass puzzle of $\Lambda_c(2940)^+$. This fact shows that $\Lambda_c(2940)^+$ can be assigned as a $\Lambda_c(2P,3/2^-)$ state when considering the unquenched effect from the $D^*N$ channel. Besides giving a quantitative calculation of the mass, we also further provide a semi-quantitative analysis to explain why $\Lambda_c(2940)^+$ under the
$\Lambda_c(2P,3/2^-)$ assignment has a narrow width. By this two steps, the nature of $\Lambda_c(2940)^+$ has been shed light on under unquenched picture.

How to test this unquenched scenario for  $\Lambda_c(2940)^+$ will be a crucial criterion we have to face. Thus, in this letter we further study its partner, a $\Lambda_c(2P,1/2^-)$ state by the same approach. We find that the mass of the discussed $\Lambda_c(2P,1/2^-)$ state is still above the $D^*N$ threshold, which means that the mass of $\Lambda_c(2P,1/2^-)$ is larger than that of $\Lambda_c(2P,3/2^-)$ in this unquenched picture, which never happen before for these observed heavy baryons and is totally different from the behavior of $\Lambda_c(2P)$ states given in the quenched picture. We strongly suggest future experiments like in LHCb and Belle II to check whether this predicted mass inversion relation for the $2P$ states of $\Lambda_c^+$ holds. If this mass  inversion relation can be established in future, this unquenched picture to $\Lambda_c(2940)^+$ will be enforced. It is a good chance for experimentalists.

Frankly speaking, the success of depicting $\Lambda_c(2940)^+$ under the unquenched scenario makes us construct a complete chain, where the
unquenched effect obviously happens in the light baryon, charmed-strange meson, charmonium, and charmed baryon families. When facing these fantastic phenomenon existing in hadron spectroscopy, we may further propose two open questions which are valuable to pay more attention to:
1) Why is the unquenched effect so significant to the $P$-wave states of hadron family? We need to further reveal inner mechanism behind this common feature.
2) Can this chain be continued? We need to check other hadron systems. In 2017, the LHCb Collaboration reported the first doubly charmed baryon $\Xi_{cc}(3621)^{++}$~\cite{Aaij:2017ueg}. It is natural to conjecture whether there exists low mass puzzle for higher doubly charmed baryon resulted by the unquenched effect, especially for the $P$-wave states.

In summary, the present study is a good start point to reveal the importance of the unquenched effect in hadron spectroscopy.
In the near future, we expect more theorists and experimentalists to join this interesting discussion.

\section*{Acknowledgments}
Si-Qiang Luo would like to thank Professor Jialun Ping for his help of Gaussian Expansion Method. This project is supported by the National Natural Science Foundation of China under Grants No. 11705072, No. 11305003, No. 11965016. XL is supported by the China National Funds for Distinguished Young Scientists under Grant No. 11825503 and the National Program for Support of Top-notch Young Professionals.

\end{document}